\documentclass[referee]{cjaa}           

\usepackage{graphicx}                   
\input{epsf.sty}                        
\input{psfig.sty}                       

\begin{document}

   \title{The early X-ray afterglows of optically bright and dark Gamma-Ray Bursts
}

   \volnopage{Vol.0 (200x) No.0, 000--000}      
   \setcounter{page}{1}          

   \author{Yi-Qing Lin
      \
      }
   \offprints{Yi-Qing Lin}                   

   \institute{Astronomy Department, Nanjing University, Nanjing 210093, China\\
             \email{yqlin@nju.edu.cn}
           }

   \date{Received~~2005 month day; accepted~~2005~~month day}
\abstract{A systematical study on the early X-ray afterglows of
both optically bright and dark gamma-ray bursts (B-GRBs and
D-GRBs) observed by Swift has been presented. Our sample includes
25 GRBs. Among them 13 are B-GRBs and 12 are D-GRBs. Our results
show that the distributions of  the X-ray afterglow fluxes
($F_{X}$), the gamma-ray fluxes ($S_{\gamma}$), and the ratio
($R_{\gamma, X}$) for both the D-GRBs and B-GRBs are similar. The
differences of these distributions for the two kinds of GRBs
should be statistical fluctuation. These results indicate that the
progenitors of the two kinds of GRBs are the same population.
Their total energy explosions are comparable. The suppression of
the optical emissions from D-GRBs should results from circumburst
but not their central engine.
   }

   \authorrunning{Lin}            
   \titlerunning{The early X-ray afterglows of optically bright and dark GRBs}  

   \maketitle

%
%
\section{Introduction}           
\label{sect:intro}
Over the 8 years since the afterglow was discovered, more than one
hundred of bursts were well-localized and their counterparts in
the X-ray, optical/IR, and radio bands were detected. About ninety
percent of these well-localized bursts are X-ray afterglow
detected, but about half of them have not optical transient (OT)
detection, which are the so called optically dark GRBs (Groot et
al. 1998; Fynbo et al. 2001; Reichart \& Yost 2001). Before Swift
era being due to the lack of early afterglow observations, it is
thought that dark bursts might be a bias of late and shallow
observations. However, very tight limits at very early phases made
by Swift UV-optical telescope (UVOT) have shown that about $50\%$
of swift GRBs are indeed phenomenally dark (Roming et al. 2005).
The nature of the dark GRBs becomes a great issue. Several
arguments have been proposed for explanation of the nature of dark
bursts. Extinction by dust and gas of host galaxy (e.g., Taylor et
al. 1998; Djorgovski et al. 2001; Piro et al. 2002) and/or
circumburst absorption (Lazzati, Covino, \& Ghisellini 2002; Fynbo
et al. 2002) are intuitionistic explanations. However, the
faintness and relatively rapid decay of the afterglow of bright
GRB 020124, combined with the low inferred extinction, indicate
that some dark bursts are intrinsically dim and not dust obscured
(Berger et al. 2002). The Ly-$\alpha$ blanketing and absorption
effect due to high redshift is also proposed (Fynbo et al. 2002;
Groot et al. 1998). However, the redshifts of two typical dark
bursts, GRB970828 and GRB000210, are normal as bright GRBs.
Recently, Roming et al. (2005) argued that dark GRBs may be
intrinsically faint and/or high efficiency gamma-ray emissions,
which should result in their cooling frequency closed to the X-ray
band (Pedersen et al. 2005) and faint at optical wavelengths (e.g.
Lazzati, Covino, \& Ghisellini 2002; Fynbo et al. 2002).

X-ray afterglow is a main probe to detect the difference of bright
and dark GRBs. De Pasquale et al. (2003) systematically compared
the X-ray fluxes by extrapolating the X-ray flux to 10 hours after
GRB trigger and found that dark GRBs tend to have a lower X-ray
fluxes. Jakobsson et al. (2004) used a jointed optical-to-X-ray
spectral index to discriminate the dark and bright GRBs by the
X-ray and optical afterglows at $11$ hours since GRB trigger. Rol
et al. (2005) try to quantify the degree of the optical darkness
by comparing optical upper limits and the inferred optical fluxes
from X-ray fluxes based on standard afterglow model. However, two
significant biases are involved in the late X-ray afterglow data
used by previous authors. The first one is sample biased. Being
due to the lack of early and deep optical observation, some
previous dark GRBs might be bright GRBs. The optical afterglow
observations of previous GRBs were made at significantly different
epoch. This also results in an inhomogenous effect for the sample
selection. Secondly, XRT observations have revealed that the early
X-ray afterglows of GRBs are enormously different from the late
ones. In this work we systematically analyze the early X-ray
afterglows observed by the Swift/XRT for bright and dark GRBs. We
collect the Swift GRB data up to June, 2005. There are 25 bursts
are included. We present our sample in section 2. The results are
presented in section 3, 4 and section 5. Conclusion and discussion
are presented in section 6.
\section{Samples}
\label{sect:Obs}
For seeking of homogenity and reliability, we include only Swift
GRBs into our sample. Twenty-five GRBs are included. We identify
those GRBs without OT detection by Swift/UVOT and(or) ground-based
telescopes as dark GRBs. In our sample 12 bursts are dark GRBs.

The X-ray afterglow of the bursts in our sample are observed by
Swift/XRT from $\sim 10^2$ s up to $10^5$ s since GRB trigger. We
measure the X-ray afterglows at a given time for our purpose. This
given time should be early enough and the X-ray fluxes at this
time should be reliably measured from the XRT light curves of most
bursts. We take this time as 1 hour after GRB trigger. Our
considerations are as follows. First, most of the XRT light curves
have a bright and steep tail in the early phase lasting from $\sim
10^2$ s up to $\sim 10^3$ s. These tails are believed to be from
prompt emissions. To reduce the contamination from the tail
emissions, we should select a time that it is later than $10^3$
seconds. Second, more than half of the XRT light curves have a gap
around 1500-3000 seconds lacking of observations. We should also
skip this period. We notice that around 1 hour since GRB trigger
most of XRT light curves begin to evolve as power law with a
normal index ($\sim -1$). At this time the fluxes are also not
affect by the jet effect\footnote{The jet break is usually greater
than half a day}. We thus study the X-ray flux at $1$ hour since
GRBs trigger.

Their X-ray afterglow fluxes ($F_{X}$) at 1 hour  after GRB
trigger are read off or extrapolated/interpolated from their X-ray
light curves observed by Swift X-ray telescope (XRT). Their
gamma-ray fluences $S_{\gamma}$ and the duration ($T_{90}$) in
15-350 keV are also collected from literature. They are listed in
Table 1 with the followings headings: GRB, gamma-ray fluence
($S_\gamma$) in 15-350 keV band (in unit of $10^{-6}$ ergs
cm$^{-2}$), GRB duration ($T_{90}$) in 15-350 keV, X-ray afterglow
flux ($F_{X}$) in 0.3-10 keV band at 1 hour since GRB trigger, and
references.


\section{Early X-ray Flux as a Function of Gamma-ray Fluences}
With the data shown in Table 1 we show the two-dimensional
distributions of the gamma-ray fluences and the X-ray fluxes in
Figure 1. It shows that the two quantities are correlated for both
the B-GRBs and the D-GRBs (panel a), with Spearman correlation
coefficient $r=0.79\pm 0.40$ (the chance probability $p<0.0001$)
for B-GRBs and $r=0.60\pm 0.58$ ($p=0.01$) for the D-GRBs. The
best fitting results are also shown in the panel (a) of Figure 1.
They show that the D-GRBs tend to have a larger ratio of
$S_{\gamma}/F_X$ than the B-GRBs. The dispersion of the
correlation for the D-GRBs is significantly larger than that for
the B-GRBs.

From Figure 1 one can observe that both $S_{\gamma}$ and $F_X$
expand almost the same ranges for the D-GRBs and B-GRBs. While the
$S_\gamma$ of the D-GRBs tends to be slightly larger than that of
the B-GRBs, the $F_X$ of the D-GRBs tends to be slightly smaller
than that of the B-GRBs.

\begin{figure}
   \vspace{2mm}
   \begin{center}
   \hspace{3mm}\psfig{figure=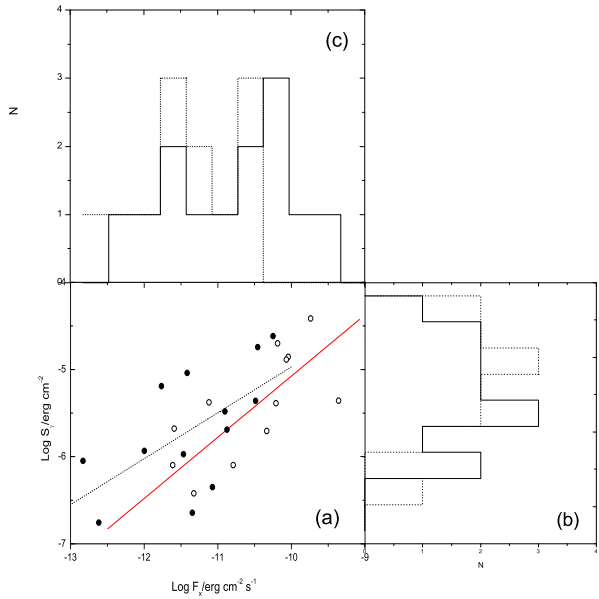,width=80mm,height=80mm,angle=0.0}
   \caption{Two-dimensional distributions of the gamma-ray
fluences and the X-ray fluxes for the optically bright [open
circles in panel (a) and solid lines in panel (b) and (c)] and the
optically dark GRBs [solid circles in panel (a) and dotted lines
in panels (b) and (c)]. The solid and dotted lines in the panel
(a) are the best fit results for the bright and dark GRBs,
respectively.}
   \label{Fig1}
   \end{center}
\end{figure}

\section{Ratio of Early X-ray afterglow flux to Average gamma-ray flux}
GRBs are from cosmological distance. The observables must be
affected by the cosmological effect. Since most of the bursts in
our sample have no redshift measurements, we could not make the
cosmological corrections. The hardness ratio between two observed
energy bands is independent of the cosmological effect. We thus
study the hardness ratios of the two kinds of GRBs. The hardness
ratio is calculated by the average gamma-ray flux to the X-ray
flux, which is $R_{\gamma, X}=\overline{F_\gamma}/F_X$, where
$\overline{F_{\gamma}}$ is the average gamma-ray flux over the
duration ($T_{90}$) in 15-350 keV band . The distributions of
$R_{\gamma, X}$ for the two kinds of GRBs are shown in Figure 2.
The two-dimensional distributions in the panel (a) of Figure 2
show that two kinds of GRBs are mixed together without any
classification signatures. The panel (b) of Figure 2 shows that
the $R_{\gamma, X}$ distributions for the two kinds of GRBs are
similar, with the $R_{\gamma, X}$ of the D-GRBs being slightly
larger than that of the B-GRBs. We perform a K-S test to examine
whether or not the two distributions are from the same parent. The
significant level for the null hypothesis that two data sets are
from the same distribution is $P_{KS}=0.098$. The null hypothesis
is marginally accepted.

\begin{figure}
   \vspace{2mm}
   \begin{center}
   \hspace{3mm}\psfig{figure=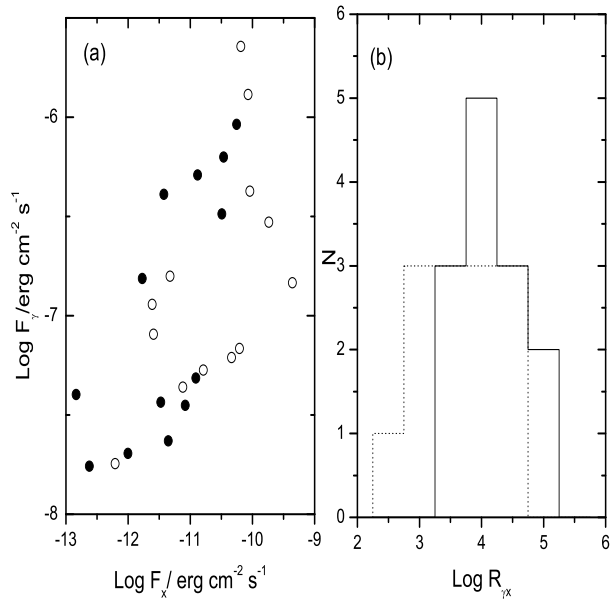,width=80mm,height=80mm,angle=0.0}
   \caption{Two-dimensional distributions of the average gamma-ray
fluxes and the X-ray fluxes for the optically bright [open circles
in panel (a) and solid lines in panel (b)] and the optically dark
GRBs [solid circles in panel (a) and dotted lines in panels (b)].
}
   \label{Fig1}
   \end{center}
\end{figure}

\section{Bootstrap test}
De Pasquale et al. (2003) found that the extrapolated X-ray
afterglow fluxes at 11 hours since GRB trigger of the D-GRBs tend
to be weaker than that of the B-GRBs with a factor $\sim 6$. The
means of the $F_X$ for the D-GRBs and B-GRBs in our sample are
$\overline{\log F_X}=-11.39\pm 0.82$ and $\overline{\log
F_X}=-10.66\pm 0.85$, respectively. The $\overline{\log F_X}$ of
the D-GRBs is slightly smaller than that of B-GRBs with a factor
of $\sim 5$. However, this difference is within the large error
scopes of the means, and it is not in any statistical sense. The
K-S test indicates that the $F_X$ distributions for both D-GRBs
and B-GRBs are drawn from the same parent. We use a bootstrap
method to examine if the slight difference between them is due to
the statistical fluctuation. We bootstrap $10^3$ pair samples of
D-GRBs and B-GRBs, and then calculate the $P_{KS}$ for each pair
sample. The distribution of the $P_{KS}$ is shown in Figure 3,
indicating the hypothesis that the pair samples are drawn from the
same parent is accepted at a significance level of $\sim 3
\sigma$. We also combine each pair samples as an assembled sample
and then apply KMM algorithm (Ashman et al. 1994) to examine if
the assembled sample can be classified as two unique groups. It is
found that the null hypothesis, which suggests that the assembled
sample is classified into two unique groups, is ruled out at a
significance level of $\sim 3 \sigma$.

\begin{figure}
   \vspace{2mm}
   \begin{center}
   \hspace{3mm}\psfig{figure=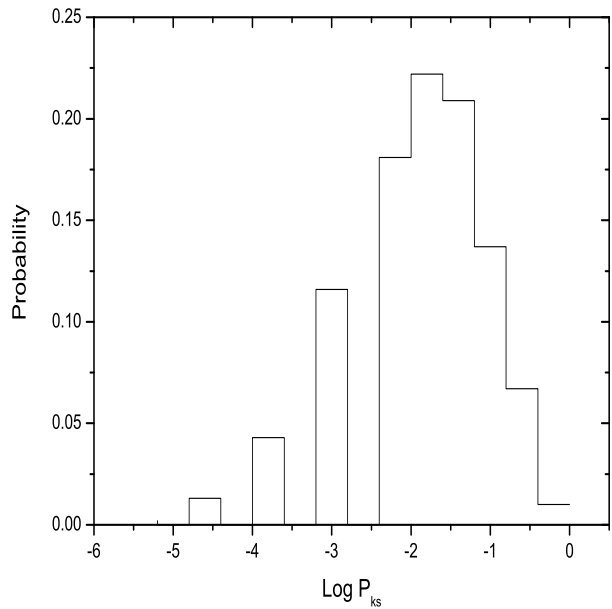,width=80mm,height=80mm,angle=0.0}
   \caption{The $P_{KS}$ distribution for $10^3$ pair bootstrap samples.}
   \label{Fig3}
   \end{center}
\end{figure}

\section{Conclusions and Discussion}
With a homogenous sample detected by Swift we have shown that the
distributions of $F_X$, $S_\gamma$, and $R_{\gamma, X}$ for both
the D-GRBs and B-GRBs are from the same parent. These results
indicate that the progenitors of the two kinds of GRBs are the
same population. Their total energy explosions are comparable. The
suppression of the optical emissions from D-GRBs should be
resulted from circumburst.

As suggested by Roming et al. (2005), the mechanisms to suppress
the optical emissions from the D-GRBs might be diverse. This
diversity may reflect the variety of the circumburst. The
extinction effect is the most popular model to explain these
D-GRBs. However, the dust in the host galaxy may be destroyed by
early radiation  from $\gamma$-ray burst and their afterglows
(Waxman \& Draine 2000; Fruchter et al. 2001). It is found that
the optical extinctions are 10$\sim$100 times smaller than
expected from X-ray absorption (Galama et al. 2001). We examine
the X-ray absorptions in our GRB sample. We do not find
systematically difference of excess $nH$ values for the D-GRBs and
B-GRBs. Extinction effect alone is hard to explain the nature of
the darkness of these GRBs. The darkness should be responsible for
more physical mechanisms. Most recently, Liang \& Zhang (2005)
found an intriguing results that within optically bright GRBs
there exists two unique classes of GRBs with late optical
afterglows. In their sample a minority of GRBs have a luminosity
dimmer than the typical ones with a factor $\sim 30$. If this is
true the nature of the dim group may cast a light on the D-GRBs.

Here we give a possible explanation that the optical dark bursts
may be caused by the synchrotron self-absorption (SSA) (Granot,
Piran 1999). If the SSA frequency is a little greater than the
observed optical frequency, which may be caused by the larger
circum-density (Sari 1998) or more loading baryons, the optical
afterglow will be darker than that in the case that the SSA can be
neglected.

\begin{acknowledgements}
YQL thanks Z. G. Dai, Y. C. Zou and E. W. Liang for helpful
suggestions and comments. This work was supported by the National
Natural Science Foundation of China (grants 10233010 and
10221001).

\end{acknowledgements}

%
\newpage
\begin{table}[]

  \caption[]{The observational data of our GRB sample}
  \label{Tab:publ-works}
  \begin{center}\begin{tabular}{ccccc}
  \hline\hline
GRB &   $S_\gamma$ & $T_{90}$ & Log($F_X$) & Ref$^ a$
              \\
 &  10$^{-6}$ ergs cm$^{-2}$  & sec&   ergs
   cm$^{-2}$sec$^{-1}$ &
\\

\hline

&&Dark GRBs&&\\

  \hline
GRB050124 & 2.10& 4.1 &-10.88&1;1;1  \\
 GRB050126  &1.10&30&-11.47&3;2;-\\
GRB050128 &  4.50&13.8&-10.49&4;3;5\\
 GRB050215b&  0.23&10&-11.35&7;;7\\
 GRB050219a &9.40&23&-11.42&7;6;7\\
 GRB050219b&  24.90&27 &-10.26&7;8;7\\
 GRB050223&0.92   &23& -12.84&7;9;7\\
  GRB050326&18.60 &   29.5 &-10.46&7;10;7\\
 GRB050410& 6.63 &43&-11.77&7;11;7\\
  GRB050421 &0.18 &10.3&-12.62&7;12;7\\
GRB050422 &1.20 &59.2&-12.00&7;16;7\\
GRB050509a  &0.46  &  13&-11.08&7;17;7\\
\hline
 &&Bright GRBs&& \\
\hline
GRB041223&38.50&130&-9.74&23;24;\\
GRB050315&4.20&96&-11.12&3;3;7\\
GRB050318&1.97&32&-10.34&7;3;7\\
GRB050319&0.80&15&-10.79&7;3;7\\
GRB050401&14.00&33&-10.04&3;25;14\\
GRB050406&0.09&5&-12.21&7;26;7\\
GRB050412&2.10&26&-11.59&7;27;7\\
GRB050416a&0.38&2.4&-11.33&7;3;7\\
GRB050502b&0.80&7&-11.61&7;28;7\\
GRB050505&4.10&60&-10.21&3;3;7\\
GRB050525a&20.00&8.8&-10.19&3;3;7\\
GRB050603&13.00&10&-10.07&7;29;7\\
GRB050607&0.89&26.5&-11.51&13;3;15\\

\noalign{\smallskip}\hline
\end{tabular}\end{center}
\end{table}

{\bf Notes:}

$^a$ In order of: $S_\gamma$ ; $T_{90}$ ; $F_X$

{\bf References:}

(1) Cummings et al. 2005a; (2) Sato et al. 2005; (3) Nousek et
 al. 2005 (4) Cumming et al. 2005b; (5) Antonelli et al. 2005; (6)
Hullinger et al. 2005a; (7) Roming et al. 2005; (8) Cumming et al.
2005c; (9) Mitani et al. 2005; (10) Cumming et al. 2005d; (11)
Fenimore et al. 2005a; (12) Sakamoto et al. 2005a; (13) Retter et
al. 2005; (14) De pasquale et al. 2005; (15) Burrows et al. 2005;
(16) Suzuki et al. 2005; (17) Hurkett et al. 2005; (23) Markwardt
et al. 2005; (24) Tueller et al. 2005; (25) Sakamoto et al. 2005b;
(26) Krimm et al. 2005; (27) Fox et al. 2005; (28) Falcone et al.
2005; (29) Fenimore et al. 2005

\label{lastpage}

\end{document}